\documentclass[a4paper,fleqn]{cas-sc}

\usepackage[numbers]{natbib}
\usepackage{graphicx} 
\graphicspath{ {figs/} }

\usepackage{natbib}

\begin{document}

\let\WriteBookmarks\relax
\def\floatpagepagefraction{1}
\def\textpagefraction{.001}

\shorttitle{MakeSbml: translating SBML and Antimony}

\shortauthors{B.E. Jardine et~al.}

\title [mode = title]{MakeSBML: A tool for converting between Antimony and SBML.}                      

%
\author[1]{Bartholomew E Jardine}

\cormark[1]


\ead{barthj@uw.edu}
\credit{Writing - Original draft preparation, Software}

\affiliation[1]{organization={Bioengineering, University of Washington},
    addressline={Box 355061}, 
    city={Seattle},
    postcode={98195}, 
    state={WA},
    country={United States}}

\author[1]{Lucian P Smith}
\credit{Software, Writing}

\author[1]{Herbert M Sauro}[%
   ]
\cormark[2]
\ead{hsauro@uw.edu}
\credit{Conceived, Software design, Writing}

\cortext[cor1]{Corresponding author}
\cortext[cor2]{Principal corresponding author}

\begin{abstract}
We describe a web-based tool, MakeSBML (\url{https://sys-bio.github.io/makesbml/}), that provides an installation-free application for creating, editing, and searching the Biomodels repository for SBML-based models.  MakeSBML is a client-based web application that translates models expressed in human-readable Antimony to the System Biology Markup Language (SBML) and vice-versa. Since MakeSBML is a web-based application it requires no installation on the user's part. Currently, MakeSBML is hosted on a GitHub page where the client-based design makes it trivial to move to other hosts. This model for software deployment also reduces maintenance costs since an active server is not required. The SBML modeling language is often used in systems biology research to describe complex biochemical networks and makes reproducing models much easier. However, SBML is designed to be computer-readable, not human-readable. We therefore employ the human-readable Antimony language to make it easy to create and edit SBML models.
\end{abstract}


\begin{highlights}
\item MakeSBML web application makes editing System Biology Markup Language (SBML) models easy with its use of Antimony modeling language.
\item MakeSBML speeds up development of system biology models by simplifying the reuse and expansion of existing SBML models.
\item Since MakeSBML is a client-side, web based application it can be hosted almost anywhere with minimal server-side resources.
\end{highlights}


\maketitle

\section{Introduction}

  To facilitate model sharing and reproducibility in the systems biology community the Systems Biology Markup Language (SBML) standard \cite{Hucka2003} is often used. However, this modeling standard can be difficult to read and edit without additional tools~\cite{Choi2018,Copasi2006,VCell2008} and, in fact, was never designed for human consumption. Instead, various SBML editors have been developed (ref, ref) to help users read and write SBML. In addition, at least two human-readable formats have been devised, SBML-shorthand (\url{https://github.com/darrenjw/sbml-sh}) and Antimony~\cite{Smith2009}. Antimony is a human-readable modeling language that greatly improves the ease of model creation and editing~\cite{Smith2009} of SBML-based models. A number of tools currently support Antimony, including Tellurium~\cite{Choi2018} and BioUML~\cite{kolpakov2022biouml}. A software development kit, libAntimony (\url{https://github.com/sys-bio/antimony}), is provided in the form of a C/C++ library and associated optional Python bindings. The software kit provides a very simple API that allows the inter-conversion between Antimony and SBML. 

  In order to make the software development kit available to web browsers we have translated the C/C++ into Web-assembly with associated JavaScript bindings. From this, we developed a client-side web application called MakeSBML \url{https://sys-bio.github.io/makesbml/} that allows a user to load, edit, and translate SBML and Antimony models from within the web browser.  Because the application is a purely client-side application, MakeSBML is hosted by a GitHub page which makes the installation robust and relatively maintenance-free.
    
\section{Materials and Methods}

\subsection{Overview}
The MakeSBML single-page web application uses a ported version of Antimony library~\cite{Smith2009} by translating the Antimony C++ library to JavaScript and WebAssembly \url{https://webassembly.org/}. This translation is done through Emscripten~\cite{zakai2011emscripten}, a compiler toolchain used to compile existing C/C++ code and make it available for use through a web browser. The WebAssembly and JavaScript code (libantimonyjs) can be found at the GitHub site: \url{https://github.com/sys-bio/libantimonyjs}. This site includes detailed information on converting the Antimony library to JavaScript and WebAssembly. The MakeSBML website contains JavaScript calls to the Antimony library which passes back and forth, as a string, the Antimony or SBML model along with any error messages.

\subsection{Usage}
\begin{figure}[ht]
\centering
\includegraphics[width=0.9\textwidth]{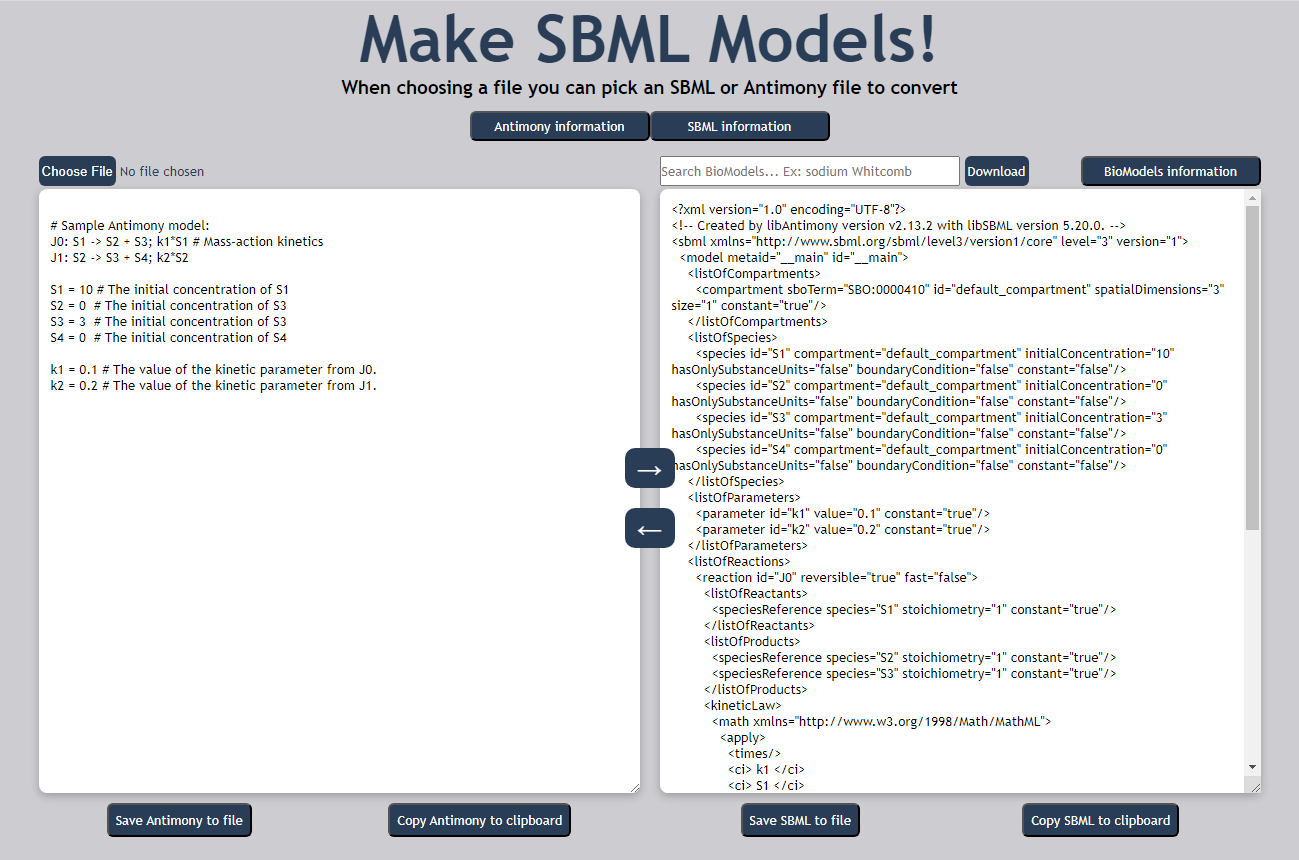}
\caption{MakeSBML web application. Figure showing the Antimony model on the left and a portion of the translated SBML model on the right. The text search box above translated SBML allows searching of BioModels model database (https://www.ebi.ac.uk/biomodels/)~\cite{BioModels2020} for model download and translation to Antimony. Note the verbosity of the SBML model compared to the antimony model.}
\label{fig:screen}
\end{figure}

Using the MakeSBML web application is straightforward (Figure~\ref{fig:screen}).  Go to https://sys-bio.github.io/makesbml/ and place either an Antimony or SBML model into the appropriate text box and press the arrow corresponding to the translation needed (Figure 1). Models can be loaded from the user's computer, downloaded from the BioModels model database (https://www.ebi.ac.uk/biomodels/), pasted or typed directly into the appropriate text box. Model editing with MakesSBML makes it much simpler to create and edit SBML models. If your modeling software supports SBML but editing the model is difficult, editing with MakeSBML is often easier and quicker. Just load the SBML XML text in MakeSBML, edit the Antimony version, and save the translated SBML version for use in your simulation software of choice.  

\subsection{Implementation}
The critical part of the MakeSBML implementation is the loading of the libAntimony JavaScript module. This is done through a call to the JavaScript wrapper libantimony.js which has mappings to the WebAssembly file libantimony.wasm. For MakeSBML, loading the libAntimony module in the browser looks like this:
\begin{verbatim}   
    libantimony().then((libantimony) => {
     //	Format: libantimony.cwrap( function name, return type, input param array of types).
     loadString = libantimony.cwrap("loadString", "number", ["number"]);
     getSBMLString = libantimony.cwrap("getSBMLString", "string", ["null"]);
     getAntimonyString = libantimony.cwrap("getAntimonyString", "string", ["null"]);
     getLastError = libantimony.cwrap("getLastError", "string", ["null"]);
     jsAllocateUTF8 = (newStr) => libantimony.allocateUTF8(newStr);
     });       
\end{verbatim}
\emph{libantimony} is the name of the Antimony library and is accessed through the loading of the JavaScript wrapper contained in the file libantimony.js. The code then assigns a JavaScript variable, or alias, \emph{loadString} to the Emscripten 'wrapped' Antimony C function call, returns an integer and takes one number (pointer to a memory address containing the model string) as input. The other JavaScript aliases are \emph{getSBMLString}, which returns the SBML version of the model, \emph{getAntimonyString}, which returns the Antimony version, and \emph{getLastError}, which returns any error found in the model. The final JavaScript alias \emph{jsAllocateUTF8} is an Emscripten helper function \emph{allocateUTF8()} that allocates memory for the string and creates a pointer to it. This function was added to the Antimony JavaScript wrapper when Empscripten compiled the original Antimony code.

Below is a JavaScript excerpt using the above JavaScript aliases for passing in an Antimony model and returning the translated model in SBML:
\begin{verbatim}
  var ptrAntCode = jsAllocateUTF8(antCode);
  var sbmlResult = "None";
  var load_int = loadString(ptrAntCode);
  if (load_int > 0) {
    sbmlResult = getSBMLString();
   } else {
    var errStr = getLastError();
   }
\end{verbatim}
To pass a JavaScript string into the Antimony function \emph{loadString()} one must first call \emph{jsAllocateUTF8()} to create a pointer to the Antimony model string. \emph{loadString()} returns 0 if no errors are detected. After successfully loading the model the SBML can be retrieved by simply calling \emph{getSBMLString()}. If an SBML model is loaded Antimony translation can be obtained with the \emph{getAntimonyString()} function. See the github repository \url{https://github.com/sys-bio/makesbml} for further details of MakeSBML implementation.

\section{Discussion and Conclusion}
As modeling in systems biology becomes more complex, it is critical that researchers build upon existing work, as much time and money is invested in creating these mathematical models. The SBML standard for describing these models allows simple reproducibity of models between researchers since many simulation environments support SBML (\url{https://en.wikipedia.org/wiki/List_of_systems_biology_modeling_software}). To encourage reuse, the Antimony language provides clear text statements to describe these models allowing researchers an easier way to understand, edit, and update them for use in their own research, saving time and expense. 

In this article, we describe MakeSBML, a client-side web tool for translating between SBML and Antimony. Because we only use client-side technology, MakeSBML can be hosted by `dumb' servers such as GitHub pages. This makes it easy to port to other sites such as free resources like Google sites. Moreover, it reduces maintenance costs since an active server is not required and web technology tends to be more stable and backward compatible than equivalent desktop software. This allows tooling to continue to run long after the funding mechanism has ceased.

MakeSBML provides a quick, and simple way to edit SBML models using the Antimony modeling language. By lowering the use barrier for researchers, MakeSBML takes the process a step further by encouraging model reuse and accelerates a user's own model development and research as a whole.    

\section{Acknowledgements}

Research reported in this was supported by NIGMS and NIBIB of the National Institutes of Health under award numbers R01GM123032 and P41EB023912. Authors thank University of Washington student Tracy Chan for her assistance in the software implementation of MakeSBML.
 
Conflict of Interest: none declared.


\printcredits

\bibliographystyle{plain}
\bibliography{makeSBML.bib}

\begin{thebibliography}{1}

\bibitem{Choi2018}
Kiri Choi, J.~Kyle Medley, Matthias König, Kaylene Stocking, Lucian Smith,
  Stanley Gu, and Herbert~M. Sauro.
\newblock Tellurium: An extensible python-based modeling environment for
  systems and synthetic biology.
\newblock {\em Biosystems}, 171:74--79, 2018.

\bibitem{Copasi2006}
Stefan Hoops, Sven Sahle, Ralph Gauges, Christine Lee, Jürgen Pahle, Natalia
  Simus, Mudita Singhal, Liang Xu, Pedro Mendes, and Ursula Kummer.
\newblock {COPASI—a COmplex PAthway SImulator}.
\newblock {\em Bioinformatics}, 22(24):3067--3074, 10 2006.

\bibitem{Hucka2003}
M.~Hucka, A.~Finney, H.~M. Sauro, H.~Bolouri, J.~C. Doyle, H.~Kitano, A.~P.
  Arkin, B.~J. Bornstein, D.~Bray, A.~Cornish-Bowden, A.~A. Cuellar, S.~Dronov,
  E.~D. Gilles, M.~Ginkel, V.~Gor, I.~I. Goryanin, W.~J. Hedley, T.~C. Hodgman,
  J.~H. Hofmeyr, P.~J. Hunter, N.~S. Juty, J.~L. Kasberger, A.~Kremling,
  U.~Kummer, N.~Le~Novère, L.~M. Loew, D.~Lucio, P.~Mendes, E.~Minch, E.~D.
  Mjolsness, Y.~Nakayama, M.~R. Nelson, P.~F. Nielsen, T.~Sakurada, J.~C.
  Schaff, B.~E. Shapiro, T.~S. Shimizu, H.~D. Spence, J.~Stelling,
  K.~Takahashi, M.~Tomita, J.~Wagner, J.~Wang, and the rest of~the SBML~Forum:.
\newblock {The systems biology markup language (SBML): a medium for
  representation and exchange of biochemical network models}.
\newblock {\em Bioinformatics}, 19(4):524--531, 2003.

\bibitem{kolpakov2022biouml}
Fedor Kolpakov, Ilya Akberdin, Ilya Kiselev, Semyon Kolmykov, Yury Kondrakhin,
  Mikhail Kulyashov, Elena Kutumova, Sergey Pintus, Anna Ryabova, Ruslan
  Sharipov, et~al.
\newblock Biouml—towards a universal research platform.
\newblock {\em Nucleic Acids Research}, 50(W1):W124--W131, 2022.

\bibitem{BioModels2020}
Rahuman~S Malik-Sheriff, Mihai Glont, Tung V~N Nguyen, Krishna Tiwari,
  Matthew~G Roberts, Ashley Xavier, Manh~T Vu, Jinghao Men, Matthieu Maire,
  Sarubini Kananathan, Emma~L Fairbanks, Johannes~P Meyer, Chinmay Arankalle,
  Thawfeek~M Varusai, Vincent Knight-Schrijver, Lu~Li, Corina Dueñas-Roca,
  Gaurhari Dass, Sarah~M Keating, Young~M Park, Nicola Buso, Nicolas Rodriguez,
  Michael Hucka, and Henning Hermjakob.
\newblock {BioModels — 15 years of sharing computational models in life
  science}.
\newblock {\em Nucleic Acids Research}, 48(D1):D407--D415, 1 2020.
\newblock gkz1055.

\bibitem{VCell2008}
I.I. Moraru, J.C. Schaff, B.M. Slepchenko, M.L. Blinov, F.~Morgan,
  A.~Lakshminarayana, F.~Gao, Y.~Li, and L.M. Loew.
\newblock Virtual cell modelling and simulation software environment.
\newblock {\em IET Systems Biology}, 2:352--362(10), September 2008.

\bibitem{Smith2009}
Lucian~P. Smith, Frank~T. Bergmann, Deepak Chandran, and Herbert~M. Sauro.
\newblock {Antimony: a modular model definition language}.
\newblock {\em Bioinformatics}, 25(18):2452--2454, 07 2009.

\bibitem{zakai2011emscripten}
Alon Zakai.
\newblock Emscripten: an llvm-to-javascript compiler.
\newblock In {\em Proceedings of the ACM international conference companion on
  Object oriented programming systems languages and applications companion},
  pages 301--312, 2011.

\end{thebibliography}


\end{document}